\documentstyle[12pt]{article}

\begin{document}
{\sf 
\begin{center}
\noindent
{\Large \bf On non-Riemannian Superconductors and torsion loops}\\[3mm]
by \\[0.3cm]
{\sl L.C. Garcia de Andrade\footnote{Departamento de
F\'{\i}sica Te\'orica -- IF -- UERJ -- Rua S\~ao Francisco Xavier 524
-- Maracan\~a -- Rio de Janeiro -- RJ -- Brasil, CEP 20550--003 \\
e-mail: garcia@symb.comp.uerj.br}}  
\vspace{2cm}
\end{center}
\paragraph*{}
The geometrization of electrodynamics is obtained by performing the complex 
extension of the covariant derivative operator to include the Cartan torsion
vector and applying this derivative to the Ginzburg-Landau equation of 
superfluids and Superconductors.It is shown that the introduction of torsion makes a shift in the symmetry breaking vacuum.Torsion loops are computed from geometrical phases outside the superconductor.Inside the superconductor the torsion vanishes which represents the Meissner effect for torsion geometry.Torsion in general equals the London supercurrent.It is possible to place a limit on the size of superconductor needed to give an estimate to torsion.
\vspace{0.5cm}
\noindent
PACS number(s): 0420, 0450
\newpage
\paragraph*{}
Riemannian Superconductors have been recently investigate by Anandan \cite{1} by making use of the spontaneous symmetry breaking and conservation of energy.The strong equivalence principle for Cooper pairs was discussed.Ealier D'Auria and Regge \cite{2} have considered gravity theories with torsion and gravitationally asymptotically flat instantons.The vanishing of the gap inside a flux tube had a perfect analogy in the vanishing of the vierbein inside the gravitational instanton.Connection between torsion vortices and non-trivial topological configurations have also been pointed out in \cite{3}.In the following an argument considering the analogy between the Meissner effect and vanishing torsion in gravity is revisited.We shown that two distinct physical interpretations may be given to the Landau-Ginzburg theory with torsion.In the first torsion shifts the symmetry breaking vacuum by a torsion dependent value.In the second the Landau-Ginzburg equation is solved and a torsion potential is obtained outside the Abrikosov tube in analogy with the Maxwell field derivable from a scalar electromagnetic potential in the usual Meissner effect.Let us now consider the Landau-Ginzburg equation for superconductors
\begin{equation}
D_{\mu}D^{\mu}{\phi}=g{\phi}(|\phi|^{2}-{\lambda}^{2})
\label{1}
\end{equation}
where ${\phi}$ is the gap field and g is a coupling constant.
In the non-gravitational Meissner effect near the symmetry-breaking 
vacuum the magnitude of ${\phi}$ is constant
\begin{equation}
|{\phi}|={\lambda}
\label{2}
\end{equation}
By considering the extended covariant derivative to include 
torsion
\begin{equation}
D_{\mu}={\partial}_{\mu}-ieA_{\mu}-ifQ_{\mu}
\label{3}
\end{equation}
where $ Q_{\mu} $ are the components of the torsion vector.
Substitution of the (\ref{3}) into (\ref{1}) yields
\begin{equation}
{D^{em}}_{\mu}{D^{em}}^{\mu}{\phi}=g({|\phi|^{2}-({\lambda}^{2}+\frac{f}{g}Q^{2})){\phi}}
\label{4}
\end{equation}
Here $D^{em}$ is the electromagnetic part of the covariant 
derivative given in (\ref{3}).We also have made use of the 
following gauges $D^{em}_{\mu}Q^{\mu}=0$ and $Q^{\mu}D^{em}_{\mu}{\phi}=0$.
Thus it is easy to see from this equation that the 
symmetry-breaking vacuum is shift by a torsion energy term 
$Q^{2}=Q_{\mu}Q^{\mu}$.Situations like that may certainly appear 
in other problems in field theory like domain walls with torsion 
\cite{4,5} and problems in Superfluids \cite{6}.Let us now solve 
the equation (\ref{5}).To begin with from this equation we observe 
that the introduction of torsion shifts the symmetry breaking vacuum to the 
point $|\phi|^{2}=\frac{1}{g}({\lambda}^{2}+\frac{f}{g}Q^{2})$ Suppose that 
we work in the static case where we are from the wall and impurities and 
the general solution drifts into an asymptotic regime in which it is 
covariantly constant
\begin{equation}
D_{\mu}{\phi}=({\partial}_{\mu}-ieA_{\mu}-ifQ_{\mu}){\phi}=0
\label{5}
\end{equation}
Differentiating once again we have
\begin{equation}
{D_{\mu}D_{\nu}-D_{\nu}D_{\mu}}{\phi}=ieF_{\mu \nu}{\phi}
-if({\partial}_{\mu}Q_{\nu}-{\partial}_{\nu}Q_{\mu}){\phi}=0
\label{6}
\end{equation}
Thus the vanishing of the Maxwell tensor is not necessarily followed 
by the vanishing of the torsion.Note that equation (\ref{6}) has two 
possible solutions.In the first case assuming that the space is free of 
electromagnetic fields or $F_{\mu\nu}=0$ the vanishing of the second term 
implies a vortice term given by the vanishing of the second term implies a 
vortice term given by
\begin{equation}
{\partial}_{\mu}Q_{\nu}-{\partial}_{\nu}Q_{\mu}=0
\label{7}
\end{equation}
nevertheless in places of the superconductor where the Higgs field
${\phi}$ vanishes,torsion and electromagnetic fields do not vanish 
as happens on the inside of the Abrikosov tubes.Outside the tube 
equation (\ref{7}) is obeyed and torsion is derived from a torsion 
potential in analogy to the electromagnetic potential
\begin{equation}
A_{\mu}=-\frac{i}{e}{\partial}_{\mu}{\theta}
\label{8}
\end{equation}
where the phase ${\theta}$ appears in the solution as 
\begin{equation}
{\phi}={\lambda}e^{i\theta}
\label{9}
\end{equation}
By analogy torsion is given by
\begin{equation}
Q_{\mu}=\frac{i}{f}{\partial}_{\mu}{\alpha}
\label{10}
\end{equation}
where ${\alpha}$ is the torsion phase.To perform the above 
computations we have considered that second order terms in the 
coupling of torsion and the electromagnetic potential may be 
dropped.The second solution assumes that the space is not free of 
electromagnetic fields and equation (\ref{6}) implies that the 
electromagnetic fields can be written in terms of torsion as
\begin{equation}
F_{\mu\nu}=\frac{f}{e}(\partial_{\mu}Q_{\nu}-\partial_{\nu}Q_{\mu})
\label{11}
\end{equation}
This implies that the electromagnetic vector potential is proportional to 
the torsion vector.Similar results have been obtained by R.Hammond \cite{7} 
in the more general case of the non-Riemannian geometry with curvature and 
torsion.From definition (\ref{11}) we are able to define the following 
electric and magnetic fields in terms of torsion
\begin{equation}
E=-\frac{f}{e}({\nabla}Q_{0}+\frac{1}{c}{\partial}_{t}Q)
\label{12}
\end{equation}
and
\begin{equation} 
H=\frac{f}{e}{\nabla}XQ
\label{13}
\end{equation}
In terms of the electromagnetic field tensor $F_{\mu\nu}$ where greek 
letters here denote the four-dimensional spacetime and the latin indices
denote the three dimensional space,can be written as 
\begin{equation}
F_{0i}=-\frac{f}{ec}{\partial}_{0}Q_{i}
\label{14}
\end{equation}
and
\begin{equation}
H_{i}={\epsilon}_{ijk}F_{jk}=\frac{f}{e}{\epsilon}_{ikl}{\partial}_{k}Q_{l}
\label{15}
\end{equation}
Where we have considered the torsion gauge $Q_{0}=0$ 
to simplify matters.By considering the London second equation
\begin{equation}
F_{0i}=\frac{m}{e^{3}n_{s}}({\partial}_{0}{j^{s}}_{i})
\label{16}
\end{equation}
Substitution of equation (\ref{14}) into (\ref{16}) yields a relation 
between the electric current $j^{s}$ and torsion $Q_{i}$ as
\begin{equation}
{\partial}_{0}Q_{i}=-\frac{mc}{e^{2}n_{s}f}{\partial}_{0}(j^{s})_{i}
\label{17}
\end{equation}
Integration on both sides of equation (\ref{17}) yields the following 
relation
\begin{equation}
Q_{i}=-\frac{mc}{e^{2}n_{s}f}(j^{s})_{i}+constant
\label{18}
\end{equation}
which gives us a relation between torsion and London current.This allow us 
to determine torsion if the torsion coupling f is known or otherwise to 
determine the torsion coupling if torsion is known from other indirect 
experiment.From the mathematical point of view expression (\ref{18}) yields a natural distributional torsion when the current has a line as a support and is given by a Dirac delta distribution.In fact as we shall see bellow it may give rise to a torsion loop.Let us consider the computation of geometrical phase for the 
supercurrent $j_{s}$ like in the case Anandan \cite{8} computed for the 
cosmic strings on a torsion geometry
\begin{equation}
{\Phi}=n{\Phi}_{0}={\int}dS_{i}H_{i}=\frac{f}{e}{\int}dS_{i}{\epsilon}_{ilm}{\partial}_{l}Q_{m}=\frac{f}{e}{\int}Q_{k}dx_{k}
\label{l9}
\end{equation}
where we have applied the Stokes theorem and equation (\ref{15}).On the 
average the last equation reduces to  
\begin{equation}
Q=\frac{e}{f}{\Phi}_{0}L^{-1}
\label{20}
\end{equation}
where L is the inductance of the circuit in the superconductor.Following 
Letelier \cite{9} we shall call the above integral torsion loop.Thus 
torsion loops can be defined on the superconductors.Note that inside the 
superconductor the supercurrent vanishes and the torsion is constant.It may 
be vanish by adjusting the initial conditions.This is the torsional Meissner 
effect.Since ${\Phi}_{0}=2.10^{-7}gauss cm^{2}$ it is possible to place a 
limit on torsion as long the torsion coupling f is known.By making use of 
an estimate of torsion computed by Lammerzhal \cite{10} given by $Q=10^{-17}cm^{-1}$ 
we may compute the torsion coupling in terms of the magnetic induction of 
the superconductor as
\begin{equation}
f=L^{-1}
\label{21}
\end{equation}
Thus the knowledged of the torsion coupling f allow us to estimate and 
design a non-Riemannian superconductor for torsion detection.Another 
estimate for torsion have been recently given by Aurell \cite{11} based on 
crystal dislocations.Recently Prof.Hehl told me that the only way to couple electromagnetic fields to torsion would be in the way described in reference
(\cite{12}) where also a covariant derivative extension to Riemann-Cartan is used which is compatible with current conservation.Although our approach seems to be quite similar to them in general we arrive to some more general consequences.
\section*{Acknowledgements}
I am very much indebt to Prof.G.E.Volovik,Prof.Aurell and Prof.P.Letelier for their interest in this work and 
for sending me their reprints on the subject of this paper.Thanks are also due to UERJ and for financial support.

\end{document}